\documentclass[11pt]{article}
\usepackage{jheppub}
\usepackage{amsmath,amssymb,amsfonts,graphicx}
\usepackage{subcaption}
\usepackage{chngcntr}
\counterwithout{equation}{section}

\newcommand{\be}{\begin{equation}}
\newcommand{\ee}{\end{equation}}
\newcommand{\bea}{\begin{eqnarray}}
\newcommand{\eea}{\end{eqnarray}}
\newcommand{\beas}{\begin{eqnarray*}}
\newcommand{\eeas}{\end{eqnarray*}}
\newcommand{\ba}{\begin{array}}
\newcommand{\ea}{\end{array}}

\renewcommand*\d[2][]{%
	\mathrm{d}%
	\ifx\relax#1\relax\else
	\rule{-0.02em}{1.5ex}^{#1}\rule{0.08em}{0ex}\!
	\fi
	#2\,
}

\title{Cosmology without time-dependent scalars \\ is like quantum field theory without RG flow}

\author[]{Mark Van Raamsdonk}

\affiliation[]{Department of Physics and Astronomy, University of British Columbia,\\
6224 Agricultural Road, Vancouver, B.C.\ V6T 1Z1, Canada.}

\emailAdd{mav@phas.ubc.ca}

\abstract{Time-dependent scalar fields provide a candidate explanation for the dark energy. For these to vary on cosmological time scales, the derivative of the scalar potential in Planck units should have roughly the same magnitude as the potential itself. We emphasize that scalars with this property are present in any four-dimensional gravitational effective theory with a known UV completion via holography, provided that the dual CFT has scalar operators with dimensions of order one. Cosmological solutions without time-dependent scalars are analogous to solutions dual to the vacuum states of such CFTs while solutions with time-dependent scalars are analogous to solutions dual to the vacuum state of quantum field theories with RG flow where these CFTs provide the UV or IR fixed point. If time-dependent scalars do explain the dark energy, the gravitational effective theory describing our universe could be a $\Lambda < 0$ model associated to a holographic CFT.}

\keywords{}

\begin{document}

\notoc
\maketitle
\newpage

\section*{Introduction}

Cosmological observations are consistent with the idea that our universe is homogeneous and isotropic on large scales. Standard theoretical models take this homogeneity and isotropy of the background cosmology as a basic assumption. With this choice, the most general classical background has a metric of FLRW form, and a set of scalar fields that depend only on time. For example, in the spatially flat case, we have\footnote{We also have matter and radiation whose effects can be described by a stress-energy tensor of perfect fluid form.}
\begin{equation}
\label{FRWphi}
    ds^2 = -dt^2 + a(t) dx_i dx_i \qquad \qquad \phi_i = \phi_i(t) \; ,
\end{equation}
The standard $\Lambda$CDM model of cosmology assumes that there are no time-dependent scalar fields, though many other models have considered them (See \cite{Peebles:1987ek,Ratra:1987rm,Caldwell:1997ii} for early discussions). Including the possibility of time-dependent scalars can only improve fits to observations, though at the cost of introducing more parameters. Current observations only constrain the dark energy equation of state parameter $w$ (equal to  $(\dot{\phi}^2/2 - V)/(\dot{\phi}^2/2 + V)$ for a scalar field model with potential $V$) to within around 10 percent of the value $w=-1$ for time-independent dark energy (see \cite{Brout:2022vxf,agrawal2018cosmological} for recent discussions), so the possibility of time-varying scalars is apparently still viable.\footnote{If there are scalar fields that vary on cosmological scales, the fluctuations in these fields will correspond to nearly massless scalar particles, and these can mediate long-range forces. These particles or the corresponding forces have not been observed, but that could be because they are part of the dark sector that is belived to contain most of the matter in the universe.}

It is interesting to ask whether theoretical considerations suggest the presence or absence of time-dependent scalar fields.
From the point of view of effective field theory, there isn't much guidance. Choosing the simplest model consistent with observations suggests the pure $\Lambda$CDM model without scalars\footnote{Note however that this model may be in tension with some observations \cite{di2021realm}.}. On the other hand, including time-dependent scalars is a more generic possibility, which is furthermore already assumed to be required during the presumed inflationary phase of cosmology. 

Having scalar fields whose scale of time variation is the same order of magnitude as the age of the universe requires that the derivative of the scalar potential is similar to the scale of the potential itself: if the scalar potential varies on cosmological times scales, we have $H = \dot{a}/a \sim \dot{V}/V = V' \dot{\phi}/V$.
From the scalar evolution equation
\begin{eqnarray}
    \ddot{\phi} + 3 H \dot{\phi} + V'(\phi) = 0 \; ,
\end{eqnarray}
we have $V'/H$ as a typical scale for $\dot{\phi}$ and from the Friedmann equation we have $H \sim \sqrt{V}$ when potential energy is important (we take Planck units where $8 \pi G / 3 = 1$). Combing these, we have
\begin{equation}
\label{Vcond}
{1 \over V} \frac{d V}{d \phi} = {\cal O}(1) \; .
\end{equation} 
In other words, we want the potential to have an order one variation when the scalar field varies by order one in Planck units. It is not possible to say whether or not this is natural from the point of view of effective field theory alone.\footnote{It is sometimes suggested that having both $V$ and $V'$ sufficiently small in Planck units represents two separate fine tunings in the effective theory, making the case with time-dependent scalars less natural. However, we will see that there are natural reasons why these would represent the same scale.}

One might hope that the likely presence or absence of time-dependent scalar fields in cosmology would be informed by quantum gravity considerations. It is expected that only certain effective field theories and perhaps only certain solutions of these effective theories are allowed in a UV complete theory (such as string theory) \cite{palti2019swampland}. With complete knowledge of these allowed theories and solutions, we would ideally like to ask whether among the space of possible homogeneous and isotropic cosmological solutions with curvature scale in Planck units similar to our own universe, the presence of time-dependent scalars is generic or rare (e.g. requiring some fine tuning).

At present, we are far from having a complete understanding of cosmological solutions in quantum gravity, so we won't be able to make any direct conclusions about the likelihood of time-dependent scalars. However, we will argue in this note that the holographic approach to quantum gravity (the AdS/CFT correspondence \cite{Maldacena:1997re}) - probably our most powerful tool for understanding UV complete quantum gravitational physics - does have something to say about the question. We will suggest that for gravitational effective theories with UV completion via a dual field theory, it is natural to have scalar fields satisfying condition (\ref{Vcond}). While we don't have a microscopic understanding of cosmological solutions of these effective theories, a very similar class of the solutions (making $t$ spacelike and one of the $x$ directions timelike in (\ref{FRWphi})) are well understood and correspond to the vacuum physics of quantum field theories. Solutions without time-varying scalars correspond to the vacuum of a dual CFT associated with our effective gravity theory, while solutions with time-varying scalars correspond to quantum field theories with renormalization group flow such that the IR or UV is governed by this CFT. The latter are more generic provided that the CFTs in question have scalar operators with dimensions of order one.

\paragraph{The AdS/CFT correspondence}
The AdS/CFT correspondence allows us to make some general statements about a very large class of gravitational effective field theories without knowing explicitly what these are. According to the correspondence, any four-dimensional effective gravitational theory with a UV completion and a stable AdS solution should correspond to a dual three-dimensional CFT, whose correlation functions are defined by asymptotic gravitational observables.\footnote{This class of theories has a scalar potential with a negative extremum $-V_0$ for $\phi_i = 0$ and obeys $d^2 V / d \phi^2 \ge - 9/4 V_0$  for each scalar in order to ensure stability of the AdS solution.}${}^{,}$\footnote{In known examples, the UV completion of the effective gravitational theory involves string theory or M theory, but we do not need to assume this.} The vacuum state of this conformal field theory corresponds to the solution of the effective  gravitational theory with pure anti-de-Sitter geometry (supported by the negative scalar potential at the extremum) and vanishing scalars.

Such four-dimensional gravitational effective theories with a dual three-dimensional CFT are arguably the only ones which we can at present be sure exist as full-fledged quantum theories. Thus, it makes sense to ask our question about the genericity of scalars with potential satisfying condition (\ref{Vcond}) in the context of these theories.

In a consistent gravitational effective theory with a dual CFT, the scalar fields are in one-to-one correspondence with the scalar operators in the dual CFT \cite{Aharony1999}. The various scalar fields correspond to the various independent directions that we can move away from the extremum of the potential. The $m^2$ of each scalar field determines the scale associated with the derivatives of the potential as we move away from the extremum in the direction corresponding to that scalar.
The mass of a scalar field (working in Planck units) is related to the dimension $\Delta$ of the corresponding scalar operator in the dual CFT by
\begin{equation}
m^2/|V_0| = (\Delta^2 - 3 \Delta) \; .
\end{equation}
From this we see the key point: for a potential $V = V_0 + {1 \over 2} m^2 \phi^2 + \dots$ (ignoring higher order terms), V will change by an order one fraction when $\phi$ changes by order one in Planck units provided that the scalar $\phi$ is associated to a CFT operator whose dimension is of order one.  Thus, from the CFT point of view, having scalars for which the scale associated with derivatives of the potential is the same as the cosmological scale (our condition (\ref{Vcond})) is completely natural. For example, $\Delta = 3$ represents the boundary between relevant and irrelevant operators in a three-dimensional CFT. A complete absence of operators with order one dimension would mean that the theory has no possible relevant deformations and no irrelevant deformations except by very large dimension operators. We are not aware of any CFT with these properties, though the space of three-dimensional CFTs is not very well understood. 

Of course, CFTs dual to realistic gravitational effective field theories are special and it may be that among this special class of theories, it is more common not to have any low-dimension scalar operators. Indeed, having relatively few  operators with dimensions of order one (i.e. having a ``gap'' in the spectrum) is a hallmark of holographic CFTs required to have a dual gravitational effective theory with relatively few light fields. But the best understood examples still do typically have some low-dimension scalar operators. For example, the AdS vacua of string theory constructed in \cite{Demirtas:2021ote,Demirtas:2021nlu} with realistically small (negative) cosmological constants and small extra dimensions have a variety of light scalars associated with possible deformations of the extra dimensions in the model.

A standard concern is that in theories without supersymmetry, the scalar masses would receive quantum corrections that bring them up to the scale of supersymmetry breaking. However, similar corrections would naively bring up the scale of vacuum energy also. The dual CFT picture relates these two scales, suggesting that if there is some bulk mechanism that keeps the cosmological constant very small, it may also keep the scalar potential flat.\footnote{Of course, it may be that non-supersymmetric large $N$ CFTs with gravity duals typically have no light scalar operators, but we are not aware of any direct evidence for this. It may also be that there are no such theories, as conjectured in \cite{Ooguri:2016pdq}. In this case, the consistent gravitational effective theories we are discussing would necessarily be supersymmetric, but this supersymmetry could be broken by the time-dependent scalars in cosmological solutions.}

In summary, in the gravitational effective field theories whose UV completions are understood, the cosmological constant scale appears to be the natural scale for the derivatives $dV/d\phi$ for the lightest scalars unless the dual CFT has absolutely no scalar operators with dimensions of order one.

\paragraph{Solutions} Even if an allowed effective field theory has scalars satisfying the condition (2), there will generally still be solutions with time-varying scalars and solutions where the scalars are fixed. Further, some solutions of the effective field theory may correspond to legitimate spacetimes in the fundamental theory, while others may not. What we would like to ask is: given the space of cosmological solutions of the effective field theory that are valid from a microscopic point of view, how generic is the presence of time-varying scalars? 

Unfortunately, while the effective field theories dual to CFTs have many cosmological solutions, we do not generally have a microscopic interpretation for these\footnote{See \cite{Maldacena:2004rf,Antonini2022} for a possible microscopic description of such cosmological solutions.}, so we can't really say anything about the space of valid cosmological solutions.
On the other hand, there is a very similar class of solutions for which we do have a good microscopic understanding. These are solutions of the same form as (\ref{FRWphi}) where we switch the signature so that the previous time direction is spatial and one of the spatial directions is timelike\footnote{This analogy was also employed recently in \cite{Baumann:2019ghk} to gain insight into inflationary cosmology. A more direct relation between such solutions and $\Lambda > 0$ cosmologies was suggested in \cite{McFadden:2009fg}.}
\begin{equation}
\label{notFRW}
    ds^2 = d\tau^2 + a(\tau) dx_\mu dx^\mu \qquad \qquad \phi_i = \phi_i(\tau) \; .
\end{equation}
Within this space of solutions, there is a large class where a microscopic description is understood. These solutions are dual to the vacuum physics of quantum field theories for which the CFT associated with our effective field theory describes either the UV or the IR physics.\footnote{In these solutions, $a(\tau)$ is exponentially growing or decaying for small or large $\tau$, so that the solution locally looks like AdS there.} 

For a given gravitational effective theory, we have a solution of this type with no scalar fields turned on and solutions with scalar fields. The solution without scalar fields corresponds to the vacuum state of the unperturbed conformal field theory while solutions with $\tau$-dependent scalars correspond to the vacuum state of some field theory with an RG flow were our CFT governs the UV or IR physics. 

Provided that the conformal field theory in question has more than one relevant operator or can arise in more than one way via RG flow from a UV theory, the  solution without varying scalars represents a measure zero subset of the space of solutions.\footnote{With only one relevant operator, the coefficient of this relevant operator is a dimensionful parameter, and all values of this parameter are physically equivalent.} In this case, the solutions with varying scalars are more generic.

\paragraph{Summary} 

We have reviewed that in the best understood gravitational effective theories, corresponding to holographic CFTs, having scalars with a potential that permits variations on cosmological time scales corresponds to having scalar operators in the dual CFT with order one dimensions. To the extent that having scalar operators of dimension one is a generic expectation for such CFTs, the condition (\ref{Vcond}) for the corresponding effective field theories seems generic rather than finely tuned.

We do not have a microscopic understanding of cosmological solutions of these effective theories, but we noted that for a very similar class of solutions, those without varying scalars holographically describe the vacuum physics of the CFT associated with the effective theory, while those with varying scalars describe the vacuum physics of quantum field theories with RG flows for which the CFT governs the UV or IR physics.  

Thus, from the point of view of holographic models of quantum gravity, having scalars which vary on cosmological time scales would appear to be a generic feature rather than a finely tuned one, unless there is some special property of CFTs with realistic gravity duals (complete absence of scalar operators with order one dimensions) for which there is so far no evidence. 

Ultimately, the existence or not of time-dependent scalars should be determined by observations. But since at present these observations still allow for models with time-dependent scalars, we suggest that such models should be taken as the most natural extension of the $\Lambda$CDM model to be considered in situations when $\Lambda$CDM appears not to align with observations (e.g. when confronting the Hubble tension).

\paragraph{Relevance of $\Lambda < 0$ models}

The reader may wonder whether it is really appropriate to attempt to gain intuition about our cosmology by thinking about gravitational theories with $\Lambda < 0$ when we live in an accelerating universe. Our motivation for considering the $\Lambda < 0$ theories was simply to find a context where can ask about the genericity of condition (2) and of cosmology-like solutions with rolling scalars with some theoretical control. If $V$ sets the typical scale for the smallest derivatives of the potential in situations where we have a complete microscopic understanding, it is plausible that this continues to be true more generally.\footnote{It is interesting that for $V > 0$, the de Sitter swampland conjecture in string theory \cite{Obied:2018sgi} suggests that the scale of $V$ gives the lowest {\it possible} scale for the derivatives of the scalar potential, though it does not say whether this scale is typical.} A separate reason that it may be relevant to consider $\Lambda < 0$ models is that if our universe does have time dependent scalars, these might well be moving towards some extremum of the potential with a negative value even though the current value is positive.\footnote{According to the de Sitter swampland conjecture in string theory \cite{Obied:2018sgi}, there are no $\Lambda > 0$ vacua , so time-dependent scalars moving towards a negative or zero value of the potential is required.} That is, if we don't assume from the start that the acceleration is caused by a cosmological constant, and we accept that scalars varying on cosmological time scales may be a generic situation in cosmology, then an effective field theory with a $\Lambda < 0$ extremum becomes a reasonable possibility for describing our observed physics.


\section*{Acknowledegements}

We would like to thank Stefano Antonini, Daniel Green, Arjun Kar, Lampros Lamprou, Juan Maldacena, Petar Simidzija, Brian Swingle, and Chris Waddell for helpful discussions. This work was performed in part at Aspen Center for Physics, which is supported by National Science Foundation grant PHY-1607611. This work is supported in part by the National Science and Engineering Research Council of Canada (NSERC) and in part by the Simons foundation via a Simons Investigator Award and the ``It From Qubit'' collaboration grant.

\bibliographystyle{jhep}
\bibliography{references}

\end{document}